\documentclass[12pt]{article}
\usepackage{graphics,epsfig}

\def\dalemb#1#2{{\vbox{\hrule height .#2pt
        \hbox{\vrule width.#2pt height#1pt \kern#1pt
                \vrule width.#2pt}
        \hrule height.#2pt}}}

\let\a=\alpha  \let\g=\gamma \let\d=\delta \let\e=\epsilon
  \let\th=\theta  \let\k=\kappa
\let\l=\lambda \let\m=\mu  \let\x=\xi \let\p=\pi 
\let\s=\sigma \let\t=\tau    
\let\vp=\varphi 
\let\w=\omega      \let\G=\Gamma \let\D=\Delta \let\Th=\Theta 
\let\X=\Xi  \let\S=\Sigma  \let\Y=\Psi
 
\let\la=\label \let\ci=\cite 
  
\def\nn{\nonumber} \def\bd{\begin{document}} \def\ed{\end{document}}
\def\ds{\documentstyle} \let\fr=\frac \let\bl=\bigl \let\br=\bigr
\let\Br=\Bigr \let\Bl=\Bigl
\let\bm=\bibitem
\let\na=\nabla
\def\tU{{\widetilde U}}
\let\pa=\partial \let\ov=\overline
\def\ie{{\it i.e.\ }}
\newcommand{\be}{\begin{equation}}
\newcommand{\ee}{\end{equation}}
\def\ba{\begin{array}}
\def\ea{\end{array}}
\def\ft#1#2{{\textstyle{{\scriptstyle #1}\over {\scriptstyle #2}}}}
\def\fft#1#2{{#1 \over #2}}
\def\F#1#2{{ F_{#1}^{(#2)} }}
\def\cF#1#2{{ {\cal F}_{#1}^{(#2)} }}

\def\R{{\bf R}}
\def\sst#1{{\scriptscriptstyle #1}}
\def\oneone{\rlap 1\mkern4mu{\rm l}}
\def\e7{E_{7(+7)}}
\def\td{\tilde}
\def\wtd{\widetilde}
\def\im{{\rm i}}
\newcommand{\ho}[1]{$\, ^{#1}$}
\newcommand{\hoch}[1]{$\, ^{#1}$}
\newcommand{\bea}{\begin{eqnarray}}
\newcommand{\eea}{\end{eqnarray}}
\newcommand{\ra}{\rightarrow}
\newcommand{\lra}{\longrightarrow}
\newcommand{\Lra}{\Leftrightarrow}
\newcommand{\ap}{\alpha^\prime}
\newcommand{\bp}{\tilde \beta^\prime}
\newcommand{\cB}{{\cal B}}
\newcommand{\cO}{{\cal O}}
\newcommand{\vecx}{\vec{x}}
\newcommand{\vecy}{\vec{y}}
\newcommand{\vecp}{\vec{p}}
\newcommand{\vecq}{\vec{q}}
\newcommand{\tr}{{\rm tr} }
\newcommand{\Tr}{{\rm Tr} }

\newcommand{\cL}{{\cal L}}
\newcommand{\cA}{{\cal A}}
\newcommand{\cD}{{\cal D}}
\def\sst#1{{\scriptscriptstyle #1}}
\def\0{{\sst{(0)}}}
\def\1{{\sst{(1)}}}
\def\2{{\sst{(2)}}}
\def\3{{\sst{(3)}}}
\def\4{{\sst{(4)}}}
\def\5{{\sst{(5)}}}
\def\6{{\sst{(6)}}}
\def\7{{\sst{(7)}}}
\def\8{{\sst{(8)}}}
\def\ve{\varepsilon}
\def\vf{\varphi}
\def\F{\Phi}
\def\wg{\wedge}

\newcommand{\auth}{AUTHORS}

\def\thb{\bar{\theta}}
\def\Thb{\bar{\Theta}}
\def\barp{\bar{p}}
\def\barq{\bar{q}}
\def\barc{\bar{c}}
\def\bard{\bar{d}}
\def\e{\epsilon}

\def \bi{\bibitem}
\def \la {\label}

\def \l {\lambda}
\def\foot{\footnote}
\def \tl  {{\tilde \l}}
\def \sql {{\sqrt \l}}
\def \adss {$AdS_5 \times S^5$\ }
\newcommand{\rf}[1]{(\ref{#1})}
\def \ov {\over}

\def\th{\theta}
\def\Th{\Theta}
\def\vth{\vartheta}
\def\btheta{{\bar\theta}}
\def\ttheta{{{\tilde\theta}}}
\def\bttheta{{{\bar\ttheta}}}
\def\vth{\vartheta}

\def\ra{\rightarrow}
\def\N{{\cal N}}
\def\F{{\cal F}}
\def\uM{\underline{M}}
\def\uN{\underline{N}}
\def\uP{\underline{P}}
\def\cc{\circ}
\def\eqv{\equiv}

\def\ni{\noindent}
\def \ha{{1\ov 2}}

\def\r{{\rm r}}

\def\Y{{\rm Y}}
\def\X{{\rm X}}
\def\tY{\tilde{\rm Y}}
\def\tX{\tilde{\rm X}}
\def\dY{\dot{\rm Y}}
\def\dX{\dot{\rm X}}

\def \J {\mathcal{J}}
\def \del {\partial}

\def\dF{\dot{F}}
\def\dG{\dot{G}}
\def\df{\dot{f}}
\def \E {{\cal E}}
\def \S {{\cal S}}
\def \J {{\cal J}}

\def\ms{\mathcal{S}}
\def\mj{\mathcal{J}}
\def\soj{\fr{\ms}{\mj}}
\def \R {{\bf R}}
\def \om {\omega}
\def \bE {\bar E}
\def \x {{\cal X}}

\begin{document}
\overfullrule=0pt
\parskip=2pt
\parindent=12pt
\headheight=0in \headsep=0in \topmargin=0in
\oddsidemargin=0in

\vspace{ -3cm}
\thispagestyle{empty}

\begin{center}

{\Large\bf
Semiclassical circular strings in $AdS_5$\\
and ``long''
 gauge field strength operators
 \\
 \vspace{0.1cm}
 }

 \vspace{.5cm} { I.Y. Park\footnote{ipark@mps.ohio-state.edu },
 A. Tirziu\footnote{tirziu@mps.ohio-state.edu}
 and A.A.
 Tseytlin\footnote{Also at Imperial College London
 and  Lebedev  Institute, Moscow.
 }}\\
 \vskip 0.3cm

{Department of Physics, The Ohio State University,\\
Columbus, OH 43210, USA   }

\end{center}

\def \bi{\bibitem}
\def \la {\label}

\def \l {\lambda}
\def\foot{\footnote}
\def \tl  {{\tilde \l}}
\def \sql {{\sqrt \l}}
\def \adss {$AdS_5 \times S^5$\ }

\def \D {\Delta}
\def \thet {\theta}
 \def \t {\tau}
 \def \p {\phi}
 \def \r {\rho}
 \def \rN {{\rm N}}

\def \ov {\over}

\def \varpi {{\rm w}}

 \vspace{0.1cm}

 \begin{abstract}
We consider  circular strings  rotating with  equal spins
 $S_1=S_2=S$ in two orthogonal planes  in $AdS_5$  and suggest
that they may be dual to ``long''  gauge theory operators built out of self-dual
components of gauge field strength. As was found in
hep-th/0404187, the one-loop anomalous dimensions of the such
gauge-theory operators  are described by an anti-ferromagnetic XXX$_1$ spin chain and
 scale linearly with length $L \gg 1$. We find  that in the case of rigid
rotating string both the   classical energy $E_0$
and the 1-loop string correction $E_1$   depend linearly on the  spin $S$
  (within the stability region of the solution). This  supports
  the  identification
 of the rigid rotating string with the gauge-theory operator
 corresponding to the maximal-spin
 (ferromagnetic)  state of the XXX$_1$ spin chain.
 The energy of more  general rotating and pulsating strings  also
 happens to  scale linearly with both the
spin  and the oscillation number. Such solutions     should be dual to other
 lower-spin  states of the  spin chain,  with the anti-ferromagnetic
 ground state  presumably corresponding to   the string pulsating in two
  planes with no rotation.
\end{abstract}
\newpage

\def \OO {{\cal O}}

\setcounter{equation}{0}
\setcounter{footnote}{0}
\setcounter{section}{0}

\renewcommand{\theequation}{1.\arabic{equation}}
 \setcounter{equation}{0}

\section{Introduction}

The study of AdS/CFT duality  between string theory  in \adss
and $\N=4$ SYM theory   uncovered
a remarkable connection between
  dimensions  of  ``long''
gauge-theory operators  and  energies   of semiclassical
 strings in $AdS_5$ \ci{bmn,gkp,ft1,ft2,bmsz,kru,kt,kmmz,mik12}
 (see  also \ci{rev,beis,zare,swan} for reviews and references).
 It is important to investigate other
   examples of this connection
 and draw  lessons that may apply to less supersymmetric
 gauge theories, including pure  YM theory.

In general, quantum  operator  dimensions
and energies of quantum string states are non-trivial
functions of the 't Hooft coupling $\l= g^2_{\rm YM} N$
(or square of string tension)  and
quantum numbers $Q$ parametrizing them,
$\D=E= E(\l,Q)$. In the  limit of large $Q$
the function $E$  should
have a  gauge-theory perturbative expansion
at
small $\l$, \
 $E_{\l \ll 1} = a_0(Q) + a_1 (Q) \l +
a_2 (Q) \l^2 +...$,
and string-theory perturbative expansion at large $\l$
(the semiclassical string expansion corresponds to
${Q\ov \sql} $ being fixed),
$E_{\l \gg 1} = \sqrt \l  b_0({Q\ov \sql})
 + b_1 ({Q\ov \sql})
 +  { 1 \ov (\sql)^2} b_2 ( {Q\ov \sql} )  +...$.

In the case of ``locally BPS''  operators
dual to ``fast'' strings carrying at least one large
angular momentum in $S^5$ the structure of the two
expansions happens to be essentially the same,    with the
leading (order $\l$ and $\l^2$)  coefficients  matching precisely.
The  general pattern, however, should be an interpolation in $\l$:
terms of different order in large $Q$
expansion  may have coefficients which are non-trivial
 functions
of $\l$ and which may
have different  large and small $\l$ limits.

\bigskip

Here we will be interested  in the case of the
 semiclassical strings
moving only within $AdS_5$   which  should thus  be dual to the
gauge-theory operators  which  do not carry
$SO(6)$ R-charges.
This sector of states is very interesting
 since at least part of it
should be common to any  gauge theory.
Then the  ``long'' operators
in this sector may be
dual to  semiclassical strings  in $AdS_5$,
with anomalous dimensions having the same
 structure as string energies.

The  obvious  global charges are the
two  spins $(S_1,S_2)$, i.e. Cartans of $SO(4)$
part of the $SO(2,4)$ isometry of $AdS_5$.
A generic operator can be
represented symbolically  as a
combination of different orderings of factors in
 $\Tr ( D^{S_1}_+ D^{S_2}_* F^M)$ \ci{ft2}
where $D_+= D_1+ i D_2$, $D_*= D_3+ i D_4$ are  covariant
derivatives in gauge theory on $R^4$
and $F^M$ stands  for products of  gauge field strength
 components, scalars or  spinors
 (in general, field strength components and spinors may also
 carry the spins $(S_1,S_2)$).
One may expect that in the  semiclassical limit
of large quantum numbers
 (like  the spins  or string oscillation numbers
 related to $M$)
  the form of the dependence of $\D=E$ on
 quantum numbers  may  happen to  be the same in the small
 $\l$  (perturbative gauge theory) and the large $\l$
 (perturbative string theory) limits.

\bigskip

 The basic  example \ci{gkp}
 is provided by the folded string which is spinning in one plane
  with its center at rest
in the middle of $AdS_5$  \ci{veg},
i.e. having $t=\k \tau$, $\r=\r(\s)$, $\thet= {\pi \ov 2},
\ \p_1= w \tau,$ where
 the metric of $AdS_5$ is chosen as
\be \la{met}
ds^2 = - \cosh^2 \r \  dt^2 + d \r^2  + \sinh^2\r\  (d \thet^2
+ \sin^2 \thet\ d \p_1^2 + \cos^2 \thet\ d \p_2^2)\ . \ee
Here    $S_1=S, \ S_2=0$  and the string energy at large $S$
is given  by \ci{gkp,ft1}
\be\la{loga}
E= S +  f(\l) \ln S + ...\ , \ \ \ \ \ \ \
f(\l)= a_0 \sql + c_1 + { 1 \ov \sql} c_2 + ...\ . \ee
The same form of $S$-dependence  is found for the
perturbative gauge-theory anomalous
dimension  of the twist-2 ($M=2$) operators of the structure
$\Tr ( F D_+^S F)$
where $f(\l)= a_1 \l + a_2\l^2   + ...$
(see \ci{lip,km} and references  there).

A similar   expression $E= S +   h(\l,n) \ln {S\ov n} + ...$
is found \ci{KRU}
for a closed
    rotating  string with $n$ spikes
which should be dual  to an operator containing
$n$ fields and $S\gg n $ covariant derivatives, i.e.
$\OO \sim \Tr ( D^{s_1}_+ F ...D^{s_n}_+ F)+ ...$, \
$S= \sum_{i=1}^n s_i$ (see also \ci{gor,bgk}).\foot{In
 this case one is able
to establish  \ci{KRU}  a
qualitative map between coherent operators and semiclassical strings
by
matching an action for the spikes motion
with a coherent state  effective action  following from the 1-loop
gauge theory dilatation operator.}
The emerging picture is that the
  spikes (or singularities in $\rho(\s)$)
should correspond
to field components  in $\OO$ while the spin should be
 represented by the total power of the covariant derivative $D_+$.

\bigskip\bigskip

An  important  question is  about the structure
of operators dual to other semiclassical string  states
in $AdS_5$, e.g., pulsating circular strings \ci{minna}
(see also \ci{smed})
or rigid  circular strings rotating simultaneously in two
orthogonal planes
\ci{ft2,art}.
Here we shall argue  that the  circular
rotating and pulsating strings in $AdS_5$ which
 have  $S_1=S_2$
should be dual to  gauge-theory operators built out of the
self-dual components  of the gauge field strength \ci{fhz}, i.e.
$\Tr ( F^{(+)}_{m_1 n_1}... F^{(+)}_{m_L n_L})$.\foot{The
$SO(4)$ representations may be labeled
as  $(S_L,S_R)$  where $S_L= \ha(S_1+S_2),\  S_R= \ha(S_1-S_2)$.
The self-dual field strength  described by $(1,0)$ representation then
corresponds to $S_1=S_2$.}
 Operators with  extra covariant derivatives
should  probably   correspond to more general
 rotating and oscillating string configurations
 which are no longer circular (i.e. have $\r$ depending on $\s$).

The sector of  self-dual  field-strength operators  was studied   in
 \ci{fhz,bfhz}
 where it was found  to be closed under renormalization
 to 1-loop order and  the corresponding 1-loop anomalous dimension
 matrix   was  identified with the Hamiltonian of an  integrable
 anti-ferromagnetic  XXX$_1$  spin chain.
In contrast to the case of operators with many  covariant derivatives
and few  fields whose anomalous  dimension is logarithmically
suppressed compared to their  canonical dimension   $L\sim S\gg 1$,
here the anomalous dimension  is  proportional to the
operator length itself,  i.e.  for  large $L$ one should have
\be \la{di} \D
= L + \g(\l) L + ... \ , \ \ \ \ \ \ \ \g(\l)= c_1 \l + c_2 \l^2 +
... \  . \ee
The 1-loop spin chain contains the ferromagnetic
highest-spin state \ci{fhz,bfhz} which may be represented as
$\OO_S \sim  \Tr (F_{1+i2,3+i4})^S$ or   $\Tr ([D_+,D_*])^S$;  for
this operator $L=2S$ and  in $\N=4$ SYM theory
$c_1= { 3 \ov 8\pi^2}$ \ci{bfhz}. Reducing the
spin (i.e. contracting some indices of the field strength factors)
 one gets to the lowest-energy zero-spin (Lorentz scalar)
anti-ferromagnetic
state $\OO_L$  for which $c_1={ 1 \ov 8\pi^2}$  \ci{bfhz}.

\bigskip

A natural  candidate for a semiclassical string state
dual to the ferromagnetic  $\OO_S$ operator
seems   to be the rigid circular rotating string solution of
\ci{ft2,art} which has the required
spin quantum numbers $S_1=S_2=S$.
This  string is positioned
at fixed radius $\rho=\r_0$ and has $t=\k \tau, \
\p_1= w \tau + m \s,\ \p_1= w \tau - m \s,   \ \theta = {\pi \ov 4}$.
However, as was found in \ci{ft2}, its classical energy
has rather unusual large $S$ behaviour,
$E= 2S + n_1 (\l S)^{1/3} + ..., \ \ n_1= 3\times  2^{-4/3}$,
while the  expected asymptotics of anomalous dimension
of  gauge-theory  operators  should be $S$, not $S^{1/3}$.
 It was conjectured in \ci{ft2} that perhaps the general
 interpolating expression   may  be
 $E=\D= 2S  + [ p(\l) + q(\l) S]^{1/3} + ... $,
 where  $q_{\l \gg 1} = \l ( c_0 + {c_1 \ov \sql} +... ) $
 and $q_{\l \ll 1} = \l ( b_1 + {b_2 \l} +... ) $.
 Then at small $\l$ and large  $S$ such that $\l S  \ll 1$
 we would get the expected  behaviour\foot{Note that the gauge-theory
 expansion assumes that one first
 expands in $\l$ and {\it then}  takes $S \gg 1$, so
 one should indeed assume that   $\l S  \ll 1$.}
 \be \la{gen}
 \D =   f(\l) S+ ...\  ,
 \ \ \ \ \ \ \  \ f(\l)_{_{\l \ll 1}}
 = 2 +  a_1 \l + a_2 \l^2 + ... \ . \ee
Here we  would like to  propose  a different
 way\foot{An argument against the above conjecture that
 $E=\D= 2S  + [ p(\l) + q(\l) S]^{1/3} + ... $
 is that it  suggests that  perturbative  SYM anomalous dimensions
 may scale as higher powers of the length $S$
 at higher loops. However, the linearity of the anomalous dimensions
 in the length is a very general property   which is a
consequence of the locality of  interactions in the spin chain. It
should hold  at least  until the order of perturbation
theory  comparable to the length when it may be modified by the
effects of winding contributions \ci{bds}. We are grateful
to K. Zarembo for this remark.}
to reconcile the
$E(S)$ behaviour on the string theory and the gauge theory sides,
supporting the  validity  of  \rf{gen} also at strong
coupling.

\bigskip

As was found   in \ci{ft2,art}, the circular $S_1=S_2$ solution is
unstable  if the semiclassical parameter $ \S\equiv {S\ov \sql}$
is bigger than a critical value $\S_{max}\approx 1.17$
(for winding number $m=1$). This  instability is of a
``strong'' type,
i.e. there is an infinite number of  bosonic fluctuation modes
which  are
tachyonic.\foot{In  the case of
a similar $J_1=J_2$ spinning string in $S^5$  there is only
one unstable mode \ci{ft2}.}
 This suggests  that one can trust this solution
(and, in particular, compute quantum string $\a'$ corrections
to its energy
and  try to interpolate to small $\l$ region)  only
in the stability interval  $ 0 < S < \sql \S_{max}$.
Since $\l$ is large on the string side, this interval
still includes
large values of $S$,  and thus we may hope to be able
compare to large
$S$, large $\l$   asymptotics
  of the exact anomalous dimension.

Indeed, by  analyzing the behaviour of the classical string energy
$E_0= \sql \E(\S)= \sql \E({S\ov \sql} ) $ with $\S$  we will find (in
section 2) that  $\E$  starts as $\sqrt{ 2 \S}$  at small $\S$
(as appropriate for a string in nearly  flat space)\foot{The small
value of $\S$ means  that the radial position of
the string $\r_0$ is also
small  but then the string is located
in the central region where    the $AdS_5$ curvature is small.}
and then goes as a straight line  near $\S $  of order 1,
i.e. $S \sim \sql \S_{max}$.
Moreover, this linear  behaviour of $E$ with $S$
in the region close to $ \sql \S_{max}$  is  preserved
at the string 1-loop   level: computing  the leading quantum
 correction $E_1$
 to the spinning
  string energy as in \ci{ft3,fpt,ptt} we will find that  again
   $E_1 \sim S$+const.
  This suggests  that in general,
  for  $\S$ far enough from zero  but  within the stability
  region, one should have
  \be \la{gee}
  E = f(\l) S + q(\l) + ... \ , \ee
  \be  \la{gn}
  f_{\l \gg 1} = p_0 + {p_1\ov \sql} + ...\ ,\ \ \ \ \ \ \ \
 h_{\l \gg 1} =\sql q_0 + {q_1} + ...\ .  \ \ \ee
 Our  results  for the   tree-level and
  1-loop coefficients are (for $m=1$)
  \be \la{ggg}
  p_0= 2.5 \ , \  \ \ \ \ \  p_1= -0.9 \ , \ \ \ \ \
   q_0 = 0.65 \ , \ \ \ \  q_1= -2.38  \ . \ee
  The matching of the spin quantum numbers and this
   linear dependence on $S$  thus supports
  the  identification of   the
  gauge-theory operator corresponding
  (1-loop order)
  to the ferromagnetic state of
  the XXX$_1$ spin chain   and the $S_1=S_2$
  rigid spinning string solution.

\bigskip

  What  about other $S_1=S_2$  string states
  that should be dual to  other  $\Tr ( F^{(+)}_{m_1n_1}...
  F^{(+)}_{m_L n_L} )$ operators
   built out of the
   self-dual field strength components  which have length
   $ L= 2S + \rN $  bigger  than the spin $2S$?
   It seems natural to expect that these should still be circular
   strings; however,
    reducing the spin  may cause the radial coordinate
   to change with time, i.e. such strings may   not be rigid, i.e.
  they  may  be  oscillating as well as
   rotating. The lowest 1-loop anomalous dimension
   anti-ferromagnetic state should then be dual
   to the $S_1=S_2=S\to 0$ limit of such  string, i.e.
   to the  pulsating string similar to the one
   considered in \ci{minna}
   (but now stretched not in one  but in  two orthogonal
   planes).
 As we shall find in section 3, the  classical
 energy of such  general  $S_1=S_2$  rotating and pulsating
 string solution (described by the ansatz $t=t(\tau), \
 \r=\r(\t), \ \p_1=  \vp( \t) + m \s,\ \p_1=\vp(\t) -m \s,
 \ \theta= {\pi\ov 4}$)
 is  again   a linear function of both
  the spin $S$ and the oscillation number $\rN$
  in the region where the corresponding semiclassical parameters
  $\S= {S\ov \sql}$ and $\N = { \rN\ov \sql}$ are of order 1.
  In particular, in the limit of $S\to 0$  we should  get
  $E=  h(\l) \rN + ..., \ h(\l)= h_0 + {h_1\ov \sql} + ...  $.

\bigskip

More general solutions  where $\r$ depends on $\s$
 may correspond to operators
with extra covariant derivative insertions.
The extreme case will
be a folded and bended string rotating in two planes
(see sect. 6 of \ci{afrt}
and \ci{art}).

One interesting open question is  whether the pulsating
 solution  found in the zero-spin  $S_1=S_2\to 0$ limit  is stable;
 in general, one expects that pulsation  improves stability
 (this appears to be
  the case when pulsation is taking place in $S^5$
direction  \ci{lars}). It would  be important  also  to explore
  if there is some analog
 of a ``reduced'' sigma model
  action on the string side and a coherent
  state action on the
 XXX$_1$ spin chain side that can be matched
  as was done in the examples
 of
 ``fast'' string moving in  $S^5$ \ci{kru} and rotating
 string with spikes
  in $AdS_5$ \ci{KRU}.
 In the present $S_1=S_2$
 case the spinning strings are, in general,  not ``fast''
 (cf. \ci{mik12}),
 so presumably one should not drop all time derivatives
 of the  ``transverse'' string coordinates in taking an appropriate
 semiclassical limit.
 The gauge-theory side is described by an
  {\it anti}-ferromagnetic
 spin chain for which it is indeed natural to expect a 2-d
  Lorentz-invariant coherent state action \ci{fhz}.

\bigskip\bigskip

We shall start in section 2 with a general discussion
of circular strings rotating in two planes in $AdS_5$.
In section 2.1 we shall show that the classical energy of the
 rigid (non-pulsating)
string solution of \ci{ft2}  scales  linearly  with
spin for $\S \sim 1$.  In section 2.2 we shall discuss the 1-loop
world-sheet   correction to the energy of the rigid string and
demonstrate that it also scales linearly with spin within the
stability region of the solution.

In section 3 we shall consider  other
rotating and  pulsating string solutions with $S_1=S_2$
and find that their  classical energy is also  linear
both in the  spin and the oscillation number.

In Appendix A we shall  point out   that a solution describing
circular string positioned at fixed radial distance $\r$
and oscillating in the third angle $\theta$ of $S^3$
is actually related   by a global $SO(4)$  symmetry transformation
to the rigid string solution of \ci{ft2,art}.
In Appendix B we shall present  some details
 about the derivation of the
 spectrum of quadratic fluctuations  near the rigid
rotating string which are used in computation of the 1-loop correction
to its energy in section 2.2.
Appendix C  we shall  review  the solution of \ci{minna} describing
 string pulsating in one plane of $S^3$ and find its energy
 as a function of the oscillation number $\rN$
 following the procedure used for strings pulsating in $S^5$ in
  \ci{kt}.
 In Appendix D we apply a similar approach
 to determine   the energy
 as a function of the spin and oscillation number
 for  more general  rotating and pulsating solutions
  of section 3.

\renewcommand{\theequation}{2.\arabic{equation}}
 \setcounter{equation}{0}

 \section{Circular $S_1=S_2$  rotating  string  in $AdS_5$}

We would like  to
 study a class of circular rotating and pulsating string
solutions in $AdS_{5}$
with the aim of establishing the dependence of the
string  energy on
the spin and the oscillation number.

 Written in terms of
 3 complex coordinates
\begin{equation}\la{py}
\Y_{0}\equiv Y_{5}+iY_{0}\ , \quad \Y_{1}\equiv Y_{1}+iY_{2}\ ,
\quad \Y_{2}\equiv Y_{3}+iY_{4}\ , \ \ \ \ \Y_{r}^{*}\Y^{r}=-1\ ,
\end{equation}
 \bea &&\Y_{0}=\cosh \rho \ e^{it}, \quad \Y_{1}=\sinh
\rho\  \sin \theta \;e^{i\phi_{1}}, \quad \Y_{2}=\sinh \rho\  \cos
\theta \;e^{i\phi_{2}} \ , \label{cartcoord1}
 \eea
 the $AdS_5$ metric \rf{met} is $ ds^{2}=d\Y_{r}^{*}d\Y^{r},$ where
$r,s=0,1,2$ ($\Y^{r}=\eta^{rs}\Y_{s}$, with $\eta^{rs}=(-1,1,1)$).
In this paper we will  use  conformal
gauge. The $AdS_{5}$ part of the  string action  is then
\begin{equation}
I=\sqrt{\lambda}\int d\tau\int^{2\pi}_0
\frac{d\sigma}{2\pi}\ L_{AdS} \ , \label{stringact}
\end{equation}
\begin{equation}
L_{AdS}=-\frac{1}{2}\left[-\cosh^2 \rho\  (\partial t)^2+(\partial
\rho)^2+\sinh^2 \rho\  ((\partial\theta)^2+\sin^2 \theta
(\partial\phi_{1})^2+\cos^2 \theta (\partial\phi_{2})^2)\right]
\nonumber
\end{equation}
We shall look for circular string  solutions of the form
($m_1,m_2$ are integers)
\begin{equation}\la{ana}
t=t(\tau), \quad \rho=\rho(\tau), \quad \theta=\theta(\tau), \quad
\phi_{1}=\varphi_{1}(\tau)+m_{1}\sigma, \quad
\phi_{2}=\varphi_{2}(\tau)+m_{2}\sigma\ .
\end{equation}
The integrals of motion
\begin{equation}
\mathcal{E}=\cosh^2 \rho\ \dot{t}, \quad \mathcal{S}_{1}=\sinh^2
\rho\  \sin^2 \theta \ \dot{\varphi}_{1}, \quad
\mathcal{S}_{2}=\sinh^2 \rho\ \cos^2 \theta\  \dot{\varphi}_{2}
\end{equation}
determine  the  global $SO(2,4)$ charges (energy and two spins)
\be \la{dee}
E\equiv \sqrt{\lambda}\mathcal{E}\ , \ \ \ \ \
S_{1}\equiv \sqrt{\lambda}\mathcal{S}_{1}\ , \ \ \ \ \
S_{2}\equiv \sqrt{\lambda}\mathcal{S}_{2}.  \ee
One of the two conformal gauge constraints implies
\begin{equation}
m_{1}\mathcal{S}_{1}+m_{2}\mathcal{S}_{2}=0\ .
\end{equation}
In this paper we will be interested in the subsector
of states  for which
\be
\mathcal{S}_{1}=\mathcal{S}_{2}\equiv \mathcal{S}\ , \ \ \ \ \ \
m_{1}=-m_{2}\equiv m\ .
\ee
As explained in the Introduction,
they are expected to be related to gauge-theory operators built out
of the self-dual components of the gauge field strength.
 The equations of motion for
$\rho$ and $\theta$ then are
\begin{equation}
\ddot{\rho}+\mathcal{E}^2 \frac{\sinh \rho}{\cosh^3
\rho}-(\dot{\theta}^2-m^2) \sinh \rho \cosh \rho-4 \mathcal{S}^2
\frac{\cosh \rho}{\sinh^3 \rho \sin^2 2\theta}=0  \label{ro}
\end{equation}
\begin{equation}
\ddot{\theta} +  2 \coth \r \ \dot{\theta}\dot{\rho}
-8\mathcal{S}^2\frac{\cos 2\theta}{\sinh^4 \rho \sin^3 2\theta}=0
\label{tet}
\end{equation}
They imply conservation of the  second conformal gauge constraint
\begin{equation}
\dot{\rho}^2-\frac{\mathcal{E}^2}{\cosh^2 \rho}+(\dot{\theta}^2 +
m^2) \sinh^2 \rho+\frac{4 \mathcal{S}^2}{\sinh^2 \rho \sin^2
2\theta}=0  \label{con}
\end{equation}

\subsection{Classical energy of  rigid  rotating string}

A particular solution of the above system of equations is found by
setting $\rho=\rho_{0}=$const; in this case  the string is
rotating within $S^3$ at a fixed radial distance $\rho_{0}$ from
the center of $AdS_5$. The remaining $\tau$-dependence of $\theta$
can be, in fact, eliminated by a global rotation  which  sets
$\theta=\frac{\pi}{4}$ (we  explain this in  Appendix A). The
resulting solution is then  equivalent to the rigid
 circular rotating string of  \ci{ft2}
in the $SO(4)$ rotated
 form in which it was  presented    in \cite{art}.

In the cartesian coordinates $\Y_s$  the solution
looks like \ci{art}
 ($ r_{0}\equiv \cosh \rho_{0}= \sqrt{ 1 + 2 r^2_1}$)
 \bea
 && \Y_0=r_0\,e^{i\k\t}\;,\;\;\ \ \ \ \ \ \ \
  \Y_1={r_1}\,e^{i w\t+im\s}\;,\;\;\ \ \ \ \
  \Y_2={r_1}\,e^{i w\t-im\s}\ ,  \label{crot}
 \eea
where  $t=\kappa \tau$ ($\E= \k \cosh^2 \rho_0$).
 One ends up with the following
relations
\begin{equation}
\mathcal{E}=\kappa+\frac{2 \kappa
\mathcal{S}}{\sqrt{\kappa^2+m^2}}\ ,\ \ \ \ \
 \kappa^2 \sqrt{m^2+\kappa^2}= {4 m^2 \mathcal{S}}\ ,   \ee
\be
\mathcal{S}=w r_{1}^2\ , \ \ \ \
w=\sqrt{m^2+\kappa^2}\ , \ \ \ \
 r_{1}\equiv \frac{\sinh \rho_{0}}{\sqrt{2}}= \frac{\kappa}{2m}\ .
\end{equation}
Here the relation for $w$  comes from  the equation for $\rho$ 
and for $r_1$ -- from the Virasoro condition.\foot{It is  easy to find a generalization 
of this solution 
to non-zero orbital momentum $J$ in a  big circle of $S^1$. The resulting $(E,S1,S_2;
 J)$ solution 
is then related by analytic continuation to the  circular 3-spin solution 
$(J_1,J_2,J_3; E)$  in $S^5 \times R_t$ \ci{ft2}.}

The explicit  expression for  the energy in
terms of the equal spins $\mathcal{S}$  and the
 winding number $m$   is then
 \begin{equation}
\mathcal{E}=\bigg( {m\ov \sqrt 3 A^{2/3}}  + { 2 \S m^{2/3}\ov 1+ A^{1/3}}
 \bigg) \sqrt{ A^{2/3}-A^{-1/3} + 1   } \ , \label{energy}
\end{equation}
\be A\equiv -1 +12 \frac{\mathcal{S}}{m^2}
(18 \mathcal{S}+\sqrt{324
\mathcal{S}^2-3m^2}) \ . \ee
 The energy dependence on
spin at small  $\mathcal{S}$ is the same as in flat space
\begin{equation}\la{sma}
\mathcal{E}=2\sqrt{m\mathcal{S}}+O(\mathcal{S}^{2/3})\ .
\end{equation}
 At large $\mathcal{S}$ the energy goes as \ci{ft2}
\begin{equation}\la{la}
\mathcal{E}=2\mathcal{S}+\frac{3}{4}(4m^2\mathcal{S})^{1/3}+...
\end{equation}
However,  this solution is not stable \ci{ft2} for large
spin $\mathcal{S}$ (or, equivalently, for  large $\k$ or large  $\rho_0$).
The region of stability depends on the value of $m$. For example,
for $m=1$  the solution is stable for
$\mathcal{S}\leq 1.17$ (see \ci{ft2} and Appendix B).
 The  plot of the classical energy in the
region of stability is presented in Figure 1.
We observe that while for small $\S$ the classical energy $E_0$
 goes as the square root of
$\S$ \rf{sma},
in the stability interval
$0.5 \leq \mathcal{S} \leq 1.17$ its  dependence on the spin is
approximately linear
\be \la{li}
E_0\approx \sql( 2.5\mathcal{S} + 0.65)=2.5 S + 0.65\sql
\ , \ \ \ \ \ \ \ \ \
0.5 \leq {S\ov \sql } \leq 1.17\ . \ee
As discussed in the Introduction, since in the semiclassical
approximation $S=\sql \S$ is still large,
 this supports the possibility of
identification of this classical solution
with the gauge-theory operator corresponding to the
 ferromagnetic state of the 1-loop XXX$_1$ spin chain of  \cite{fhz}.

\begin{figure}[ht]
\centerline{\includegraphics[scale=1.2]{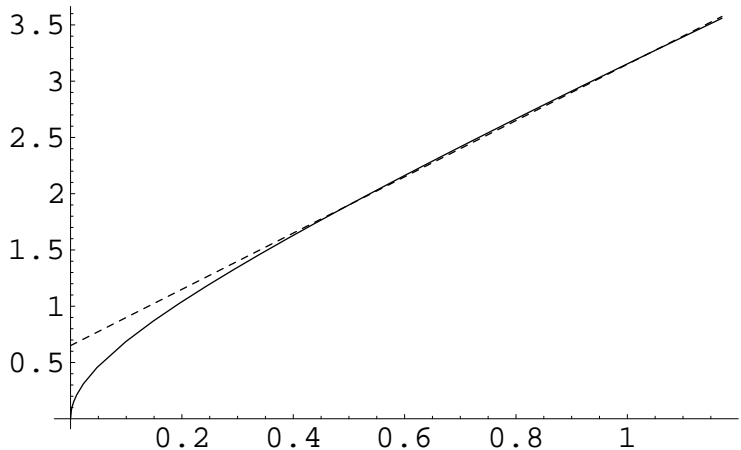}}\caption{Plot of
the classical energy $\E(\S)$
for
$m=1$.  The dashed  line is $2.5\mathcal{S}+ 0.65 $.}
\end{figure}

\subsection{One-loop correction to energy of rigid rotating
string}

To be able to extrapolate the above solution to small $\l$,
large $S$ region one is to be sure that
the $AdS_5 \times S^5$ string quantum corrections
do not modify the qualitative behaviour of the energy with spin.
Below we shall describe the result of the computation of the
1-loop correction to the energy of the rigid circular solution
following  \ci{ft2,ft3,fpt,ptt}.

The rigid circular rotating string  is a ``homogeneous''
solution for which the fluctuation Lagrangian has constant
coefficients \ci{art}.  As a result,
 the spectrum of quadratic
fluctuations can be found explicitly
(which is a  substantial  simplification compared to the case of the single-spin
folded string solution discussed in \ci{ft1}).

The three  transverse  $AdS_{5}$ bosonic fluctuations
 have characteristic frequencies determined
 by the following equation \ci{ft2,art} (see also  Appendix B)
\begin{eqnarray}\la{cu}
0&=&\omega^6-(8m^2+10 \kappa^2+3n^2)\omega^4+(16m^4+40 m^2
\kappa^2+24 \kappa^4+8 n^2 \kappa^2+3n^4)\omega^2\nn\\
&-&n^2(n^2-4m^2)(n^2-4m^2-2\kappa^2)
\end{eqnarray}
In addition,  there are two free massless ($\omega=\pm n$)  $AdS_{5}$ bosonic
modes and also five free massless $S^{5}$ bosonic
modes. The condition of stability of the solution is
that all bosonic frequencies  are  real.
That implies (we are assuming that $\S$ is positive) \ci{ft2}
\begin{equation}\la{vvv}
\mathcal{S}\leq \mathcal{S}_{max}\ , \ \ \ \ \ \ \ \
\mathcal{S}_{max}=\frac{4m+1}{8m^2}
\sqrt{(m+1)^2 - {\textstyle {1\ov 2}} }\ .
\end{equation}
In the previous subsection we have used that
$\S_{max} (m=1) = { 5 \ov 8} \sqrt{7 \sqrt 2} \approx 1.17$.
The   fermionic
frequencies  are determined (with factor of 4 degeneracy)
 by the
characteristic equation (see Appendix B)
 \bea\la{cup}
0&=& 16\,\omega^4-
  8\left( 4\,n^2 + 4\,m^2 + 5\,{\kappa }^2 \right)\,\omega^2
  +16\,m^4 - 32\,m^2\,n^2 + 16\,n^4   \nn \\
  &+&24\,m^2\,{\kappa }^2
  -8\,n^2\,{\kappa }^2 + 9\,{\kappa }^4\ ,
 \eea
with the solution
 \bea\la{huh}
 \omega^F_n= \pm \sqrt{{\ n^2 + \ m^2 + {5\ov 4}\,{\kappa }^2 \pm
 \,{\sqrt{4\,m^2\,n^2 + m^2\,{\kappa }^2 + 3\,n^2\,{\kappa }^2 +
{\kappa }^4}}}}
 \eea
The 1-loop
correction to the space-time energy $E_1$ is  given by
the sum of the characteristic frequencies (see
\ci{ft1,ft2,ft3,ptt} for details)
 \begin{eqnarray}
E_1=\fr1{\k}E_{2{\rm d}}&=&\fr{1}{2\k}\bigg[ \sum_{p=1}^{8}\left(
     | \w_{p0}^B| -|\w_{p0}^{F}|
    \right)
   +   \sum_{n=1}^\infty\;\sum_{I=1}^{16} \bigg(
    |\w_{In}^B|- |\w_{In}^{F}|\bigg) \bigg] \ .
 \label{e1}
 \end{eqnarray}
Note that here (in contrast to more subtle case
discussed in  \cite{ptt}) we can use absolute values
of the frequencies since half of the frequencies are positive and
half negative (both bosonic and fermionic characteristic
equations are expressed in terms of $\omega^2$).
It remains only to
compute the sum in (\ref{e1}). The sum is
convergent as  follows from the  large $n$
asymptotics of the non-trivial bosonic  and fermionic frequencies
\begin{equation}
\omega_{1,2 n }^{B}=|n|\pm\sqrt{3\kappa^2+4m^2}+\frac{\kappa^2}{2|n|}+O({1\ov n^2})
\ , \ \ \ \ \ \ \ \ \
\omega_{3n}^{B}=|n|+\frac{\kappa^2}{|n|}+O({1\ov n^2})\ ,
\end{equation}
\begin{equation}
\omega_{1,2n}^{F}=|n|\pm
\frac{1}{2}\sqrt{3\kappa^2+4m^2}+\frac{\kappa^2}{4|n|}+O({1\ov
n^2})
\end{equation}
Taking  into account the degeneracies and including the
contribution of the bosonic
frequencies in the $S^{5}$ directions we then get
\begin{equation}
\sum_I \omega_{In}^{B}=16 |n|+\frac{4\kappa^2}{|n|}+O({1\ov n^2})\ , \quad\
\ \ \ \ \ \
\sum_I \omega_{In}^{F}=16 |n|+\frac{4\kappa^2}{|n|}+O({1\ov n^2})\ ,
\end{equation}
which confirms the convergence of
the sum in (\ref{e1}).

Due to the rather complicated form of the explicit
solutions for the characteristic frequencies we are  unable to
 perform the sum in (\ref{e1}) analytically.
However, it is straightforward  to evaluate it numerically in
 the region of stability of the solution when the
 frequencies and thus $E_1$  are real.
 For $m=1$, the resulting one-loop  correction
$E_{1}$ is plotted  in Figure 2 as a function of
$\mathcal{S}= { S \ov \sql}.$
\begin{figure}[ht]
\centerline{\includegraphics[scale=1]{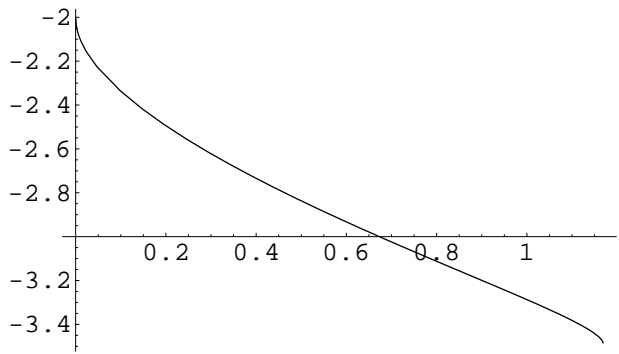}}\caption{
One-loop correction to energy $E_{1}$
as a function of  $\mathcal{S}$}
\end{figure}
We find that  there is again an
interval within the stability region
(from around $0.4$ to $1.15$) where $E_{1}$ can be well
approximated by a straight line
\begin{equation}\la{onn}
E_{1}\approx  -0.9 \frac{S}{\sqrt{\lambda}} -2.38 \ .
\end{equation}
The total energy is then $E=E_0 + E_1 + ..$, where  $E_0$
and $ E_1$ are given by
 \rf{li} and \rf{onn}. This
  supports the conjecture that the  general large-$S$
 expression for the string energy  that
 may be possible to extrapolate to weak coupling is
  given by \rf{gee}.

\renewcommand{\theequation}{3.\arabic{equation}}
 \setcounter{equation}{0}

\section{Rotating and pulsating string}

Let us now  return to the  equations (\ref{ro}),(\ref{tet}),
(\ref{con})
and look for more general
circular rotating and pulsating solutions
 with ${S}_{1}={S}_{2}$.
 Our conjecture is that they may  be dual to other (lower-spin)
 operators built out of self-dual gauge-field components.

 As was mentioned above,  having $\rho$=const while
 $\theta$ changing with time does not lead to a new solution.
 Here we shall consider the alternative case when
 $\theta$ remains fixed at $\pi \ov 4$
  while $\rho$ is allowed to change in
 time.  This will generalize the pulsating solution
 discussed in
\cite{minna}  where the string was lying only in one plane (i.e.
had  $\theta=\frac{\pi}{2}$) and  was not rotating
  (we shall review
this solution in Appendix C).

The solution with $S_1=S_2$ we are interested in
is described by  \rf{ana} with $\vp_1=\vp_2$ and fixed $\theta$,
i.e. by
the following ansatz
\begin{equation}\la{a}
t=t(\tau) \ , \ \ \
\rho=\rho(\tau), \quad\ \theta=\frac{\pi}{4}, \quad\
\phi_{1}=\varphi(\tau)+m \sigma,\ \quad \phi_{2}=\varphi(\tau)-m\sigma \ .
\end{equation}
The conserved charges are
\begin{equation}
\mathcal{E}=\cosh^2 \rho\ \dot{t}\ ,\ \ \ \ \ \ \  \quad
\S_1=\S_2=\mathcal{S}=\frac{1}{2}\sinh^2 \rho\ \dot{\varphi}\ .
\end{equation}
The conformal constraint gives the  equation of motion for
$\rho$
\begin{equation}
\dot{\rho}^2 + V(\r)=0 \ , \ \ \ \ \ \ \ \ \ V\equiv
-\frac{\mathcal{E}^2}{\cosh^2 \rho}+\frac{4 \mathcal{S}^2}{\sinh^2
\rho}+m^2 \sinh^2 \rho\ .  \label{rho2}
\end{equation}
This may  be interpreted as an equation for a particle moving in a
potential which  is growing to infinity both
at $\rho \to 0$ and $\rho \to \infty$ and having a minimum in between.
The coordinate $\rho(\t)$ thus oscillates between  a minimal amd
maximal value.

Following  the discussion of   pulsating solutions in $S^5$ in
\ci{kt}  we can  compute the oscillation  number
(which should take integer values in quantum theory) as
\begin{equation}\la{ne}
\rN=\frac{\sqrt{\lambda}}{2\pi}\oint d\rho\
p_{\rho}=\frac{\sqrt{\lambda}}{\pi}\int_{\rho_{min}}^{\rho_{max}}d\rho
\sqrt{\frac{\mathcal{E}^2}{\cosh^2
\rho}-\frac{4\mathcal{S}^2}{\sinh^2 \rho}-m^2 \sinh^2 \rho}\ .
\end{equation}
Changing the   variable to $x=\sinh \rho$ we get
\begin{equation}
\rN=\frac{\sqrt{\lambda}}{\pi}\int_{\sqrt{R}_{2}}^{\sqrt{R}_{3}}\frac{dx}{1+x^2}\sqrt{\mathcal{E}^2-\frac{4\mathcal{S}^2
(1+x^2)}{x^2}-m^2 x^2 (1+x^2)} \label{osc}
\end{equation}
where $R_{2},R_{3}$ are the two positive roots of the cubic polynomial
\begin{equation}\la{ko}
f(y)=m^2 y^2 (1+y)+y(4\mathcal{S}^2-\mathcal{E}^2)+4 \mathcal{S}^2\ ,\ \
\ \  \ \ \ \   y\equiv x^2 \ .
\end{equation}
The third root ($R_{1}$) is negative.

One can show (see Appendix D)
 that in order for this solution to exist the function
inside the square root in (\ref{osc}) must be positive. This gives
a condition on the energy, spin and the winding number $m$
\begin{equation}
\mathcal{E}^2 \geq \max \left\{4 \mathcal{S}^2,\
h(m,\mathcal{S})\right\}  \label{cond}
\end{equation}
where $h(m,\mathcal{S})$ is  the maximal root  of the
polynomial $g(z)$ (which has at least one real root)
\begin{eqnarray}\la{g}
g(z)=-4 z^3+z^2 (48 \mathcal{S}^2- m^2)-  16{S}^2 z (
12 \mathcal{S}^2  - 5m^2)
+16\mathcal{S}^2( 4\mathcal{S}^2+  m^2)^2\ .
\end{eqnarray}
The  asymptotic dependence of the energy on the oscillation  number $\rN$
and spin is worked out in Appendix D. For small $\mathcal{S}$ and
$\mathcal{N}\equiv {\rN\ov \sql}$ the energy dependence is like in
flat space
\begin{equation}\la{l}
E=2\lambda^{1/4} \sqrt{m L }+...\ , \ \ \ \ \ \ \ \   L \equiv \rN + S \ .
\end{equation}
For large $\mathcal{N}$ but small $\mathcal{S}$ we get
\begin{equation}
E=2L + d_{1} \lambda^{1/4}\sqrt{L}+  d_2
\sqrt{\lambda}  + d_3 \frac{\lambda^{3/4} }{\sqrt{L}}+O(L^{-3/2})\ ,
\label{smallS}
\end{equation}
where
\begin{equation}\la{coe}
 L\equiv \rN+S\ , \ \ \ \ \ \ \ \  d_1\approx 0.359 m^{3/2}\ , \ \ \ \
  d_2 \approx0.032 m^3\ , \ \ \
  d_3\approx-0.059 m^{5/2}
  \ .
\end{equation}
Starting with the $O(L^{-3/2})$ term, the coefficients
will depend on the spin $S$  explicitly, i.e. not only
through $L$.
For large $\mathcal{S}$ but small $\mathcal{N}$ the energy
 is given by (for $m=1$)
\begin{equation}
E=2S + v_1 \rN+ v_2 (\lambda S)^{1/3}+O(S^{-1/3})\ , \ \ \ \ \ \ \
v_1= 1.83\ , \ \  v_2= 0.45 \ .
\end{equation}
Let us now consider the dependence of $E$ on $\rN$ and $S$
in the intermediate region of $\mathcal{N}$ and $\mathcal{S}$
of order 1  (for $m=1$), where, as in the rigid string case,
the solution is
expected  to be stable (pulsation,  in general,
should improve stability).
Let us consider the region
$0.4\leq \mathcal{S}\leq1$.
According to (\ref{cond}),  this means that   $\mathcal{E}$ is
at least greater than  $2\mathcal{S}$.
For each value of the spin $\mathcal{S}$
we can determine the minimum energy $\mathcal{E}_{min}$ from
(\ref{cond}). We computed numerically the oscillation  number
$\mathcal{N}=\rN/\sqrt{\lambda}$ using (\ref{osc}) and
 plotted it  as a
function of $\mathcal{E}$ for several values of $\mathcal{S}$
from $\E=\mathcal{E}_{min}$ to $\E=5$.
Near $\mathcal{E}_{min}$ we got the expected
 parabolic  dependence ($\mathcal{N} \sim
\frac{\mathcal{E}^2}{4}-\mathcal{S}$) as in flat space since
there the
string is close to the origin of $AdS_5$.
The  string is oscillating near the radial point $\r_*$
which is not far from 0. The results are presented in the
 Table below.

\bigskip

\bigskip
\begin{center}
\begin{tabular}{|c|c|c|c|}
  \hline
  $\mathcal{S}$ & $\mathcal{E}_{min}$ & $\mathcal{N}=
  \mathcal{N}(\mathcal{E})$ & $\rho_*$\\
  \hline
  0.4 & 1.629 &  -0.727+0.401 $\mathcal{E}$ & 0.686\\
  0.6 & 2.161 &  -0.918+0.401 $\mathcal{E}$ & 0.786 \\
  0.8 & 2.665 &  -1.106+0.4006 $\mathcal{E}$ & 0.86 \\
  1 & 3.153 &  -1.292+ 0.4002 $\mathcal{E}$ & 0.919\\
  \hline
\end{tabular}
\end{center}
\bigskip

The plot of $\N(\E)$ for $\mathcal{S}=0.4$ is given  in Figure 3.

\begin{figure}[ht]
\centerline{\includegraphics[scale=0.8]{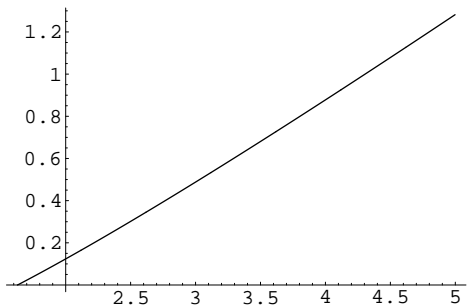}}\caption{$\mathcal{N} (\E) $
 for  $\mathcal{S}=0.4$  }
\end{figure}

We observe  that the slope of the linear function $\N(\E)$
 does not depend on
$\mathcal{S}$, i.e.
 $\mathcal{N}\approx  a(\S) +0.4
\mathcal{E}$.
For
$\mathcal{N}=0$ and $\mathcal{S}$ within the range $0.4\leq
\mathcal{S}\leq1$ we should reproduce
 the  straight line for
$\mathcal{E}=\mathcal{E}(\mathcal{S})$ found in the rigid string case.
The plot of
$a(\mathcal{S}$)  is well approximated by the
straight line  $a(\mathcal{S})=-0.248-1.017 \mathcal{S}$.
We conclude then that
\begin{equation}\la{nnn}
\mathcal{N}=-0.248-1.017 \mathcal{S}+0.4 \mathcal{E} \ ,
\end{equation}
so that the classical energy of rotating and pulsating string
has the linear form in the region where
$\rN \sim \S \sim \sql$
\begin{equation}\la{eee}
E_0 \approx 2.5 \rN+2.54 S + 0.62 \sqrt{\lambda}\ .
\end{equation}
For  $\rN=0$  this is in reasonable agreement
with the rigid string expression \rf{li}.

Written in terms of the length $L= S + \rN$
 and the spin  the energy is
 $E_0 \approx 2.5 L + 0.04 S + 0.62 \sqrt{\lambda}$,
 i.e.  grows with spin for fixed length
 as does the anomalous dimension on the gauge-theory side
 \ci{bfhz}.
 Note also that the rigid  string ($\rN=0$)
has lowest energy for given spin $S$ --
adding oscillations  increases the energy.
This is
in qualitative agreement with the gauge-theory
 interpretation where
$\Tr F^S$ operator has lowest anomalous dimension for given spin.

Repeating the above discussion
in  the  case of $\S=0$ we  find a similar linear
dependence of $E$ on $\rN \sim \sql$.
This supports the conjecture that
this pulsating string state  should be dual to
the gauge-theory operator represented
by the anti-ferromagnetic vacuum of the XXX$_1$ spin chain
 \cite{fhz,bfhz}.

\section*{Acknowledgments }
We are
grateful to  K.  Zarembo for important explanations and suggestions.
 We also thank A. Belitsky,  M. Kruczenski, A. Mikhailov and J. Minahan
 for    discussions.
This  work  was supported  by the DOE
grant DE-FG02-91ER40690. The work of A.A.T.
 was also supported by
 the INTAS contract 03-51-6346
and RS Wolfson award.

\bigskip\bigskip

\renewcommand{\theequation}{A.\arabic{equation}}
 \setcounter{equation}{0}
  \section*{Appendix A: String with  $\rho=$const  and
  oscillating $\theta$
   }

Here we would like to show that the spinning string
solution  with   constant $\rho$  which exists if  $S_1=S_2$
 and   with  $\theta$
changing with time is actually equivalent
to the solution with
$\theta={\pi\ov 4}$  we considered in section 2.2.
For $t=\k \tau$
  the
equation of motion for $\rho$ and the conformal constraint
give
\begin{equation}
\kappa^2=2m^2 \sinh^2 \rho_{0}
\ ,  \ \ \ \ \ \ \
\dot{\theta}^2 =w^2-\frac{g^2}{\sin^2 2\theta}
\end{equation}
where
\begin{equation}
w^2=m^2+\kappa^2\ ,\ \ \ \ \ \ \ \ \
 \quad g^2=\frac{16 m^4 \mathcal{S}^2}{\kappa^4}\  .
\end{equation}
Then  $\theta$ can  oscillate in the range
\begin{equation}
\theta_{min}\leq \theta \leq
\frac{\pi}{2}- \theta_{min} \equiv \theta_{max}\ , \ \ \ \ \ \ \ \
\theta_{min}\equiv \frac{1}{2}\arcsin \frac{g}{w}\ .
\end{equation}
The explicit form of the solution for $\theta$ is
\begin{equation}
\cos 2\theta=\sqrt{1-\frac{g^2}{w^2}}\ \sin 2w \tau
\end{equation}
As $\tau$ increases from zero to $w\tau=\frac{\pi}{4}$,
$\theta$ decreases from $\frac{\pi}{2}$ to
$\theta_{min}$. The
solutions for the two angles $\phi_{1},\phi_{2}$
\begin{equation}
\varphi_{1}=\arctan\big( \frac{w}{g} \tan
w\tau-\sqrt{{w^2\ov g^2}-1}\big)
\ , \ \ \ \ \ \ \
\varphi_{2}=\arctan\big( \frac{w}{g} \tan
w\tau+\sqrt{{w^2\ov g^2}-1}\big)\ .
\end{equation}
Expressing  this solution in cartesian coordinates
\rf{cartcoord1}
we get
\begin{equation}
\Y_{1}=\frac{\sinh \rho_{0}}{\sqrt{2}w}e^{i m\sigma}\left(a
e^{iw\tau}+b e^{-iw\tau}\right)\ , \ \ \
\quad \Y_{2}=\frac{\sinh
\rho_{0}}{\sqrt{2}w}e^{-i m\sigma}\left(c e^{iw\tau}+d
e^{-iw\tau}\right) \label{cart}
\end{equation}
where
\begin{equation}
a,b=\ha ( {g\pm w}-{i}\sqrt{w^2-g^2})\ , \quad\ \ \ \ \ \ \
c=\bar{a}\ , \ \ \  d=\bar{b} \ , \ \ \  ad+bc=0 \ . \ee
The corresponding oscillation  number in $\theta$ direction
(i.e. the action variable conjugate to angle $\theta$
which for periodic motion is quantized in semiclassical quantization)\foot{The string motion is described
by an integrable 1-d Neumann model \ci{afrt,art}, and the energy, spins
 and the oscillation
number  are related to its integrals of motion.}
  is
(here $p_{\theta}= \sinh^2 \rho_{0}\  \dot{\theta}$)
\begin{equation}
\rN=\frac{\sqrt{\lambda}}{2\pi}\oint d\theta\
p_{\theta}=\frac{\sqrt{\lambda}\kappa^2}{2\pi
m^2}\int_{\theta_{min}}^{\theta_{max}} d\theta
\sqrt{w^2-\frac{g^2}{\sin^2 2\theta}}
= \frac{\sqrt{\lambda}\kappa^2}{4m^2}(w-g)\ .
\end{equation}
When $w=g$ we get, as expected,   $\rN=0$:
 then  $\theta=\frac{\pi}{4}$, i.e.  there are   no oscillations.
We can also rewrite $\rN$ as
\begin{equation}
\rN=\frac{\sqrt{\lambda}\kappa^2 \sqrt{m^2+\kappa^2}}{4m^2}-S\ ,
\end{equation}
and recalling  that the energy is
\begin{equation}
\mathcal{E}=\kappa+\frac{\kappa^3}{2m^2}\ ,
\end{equation}
we conclude that it depends  on $\rN$ and $S$ only through their
sum
 $L=\rN+S$. This suggests that this pulsating
 solution is not a new one
 but a rotated version of the rigid solution with $\theta = {\pi\ov
 4}$.
 Indeed,  starting with \rf{cart}
 one can show that by an  $SO(4)$ rotation this
solution can be transformed into the rigid  rotating solution
\rf{crot} (this $SO(4)$  rotation implies also the redefinition of
the spin  $ S+\rN\rightarrow S$).

\renewcommand{\theequation}{B.\arabic{equation}}
 \setcounter{equation}{0}
  \section*{Appendix B: Fluctuations near
  rigid rotating  string
   }
Here we provide some details  for the computation
of 1-loop correction to the energy  of the rotating string solution
in section 2.2.  Let us first review the
the form of the bosonic fluctuation Lagrangian following \ci{art}.
Since the string is moving entirely in $AdS_5$
the  fluctuations in $S^5$ directions are trivial
(5 massless 2d bosons).
The fluctuation Lagrangian in the
$AdS_5$ directions near the solution \rf{crot} is
 \bea
 \tilde{L}_{AdS}=-\fr12\;\pa_a\tilde{\Y}^r\pa^a\tilde{\Y}_r^*
 -\fr12 \tilde{\Lambda}\;
  \tilde{\Y}^r\tilde{\Y}_r^*
 \ , \ \ \ \ \ \ \ \ \
 \sum_{r=0}^2\;({\Y}^r\tilde{\Y}_r^*+{\Y}_r^*\tilde{\Y}^r)=0\ .
  \label{AdSconst}
 \eea
Writing the  fluctuation fields as
\begin{equation}
\tilde{\Y}_{0}=(g_{0}+if_{0})e^{i \kappa\tau}, \quad
\tilde{\Y}_{1}=(g_{1}+if_{1})e^{i w\tau+im\sigma}, \quad
\tilde{\Y}_{2}=(g_{2}+if_{2})e^{i w\tau-im\sigma}\ ,
\end{equation}
the above constraint gives (see \rf{crot})
$
g_{0}=\frac{r_{1}(g_{1}+g_{2})}{r_{0}}
$.
Then the fluctuation Lagrangian for the five real fields
$g_{1},g_{2},f_{0},f_{1},f_{2}$ takes the form
\begin{eqnarray}
\tilde{L}&=&-\frac{1}{2}\big[\dot{f}_{0}^2-f_{0}'^{2}+4\kappa\frac{r_{1}}{r_{0}}
\dot{f}_{0}(g_{1}+g_{2})\big]-\frac{r_{1}^2}{2
r_{0}^2}\left[(\dot{g}_{1}+\dot{g}_{2})^2-(g_{1}'+g_{2}')^2\right]
\nn\\
&+&\frac{1}{2}\left(\dot{g}_{1}^2+\dot{f}_{1}^2
+\dot{g}_{2}^2+\dot{f}_{2}^2  -g_{1}'^2-f_{1}'^2
-g_{2}'^2-f_{2}'^2\right)
\nn\\
 &+&
2\left(
w g_{1}\dot{f}_{1}- m g_{1}f_{1}'
+ w
g_{2}\dot{f}_{2}+ m g_{2}f_{2}'\right)\ .
\end{eqnarray}
This determines the  bosonic characteristic
frequencies (by expanding the fields as $\sim e^{in
\sigma+i\omega\tau}$  and finding the condition of
consistency of the resulting linear system of
equations). We find that there are two massless modes
which decouple (in conformal gauge
their contribution is effectively cancelled by
the contribution of the ghosts),  and for  the remaining
coupled three directions we
obtain the characteristic equation \ci{ft2,art}
\rf{cu}, or
 $f_n(\omega^2)=0$, where
\begin{eqnarray}
f_n(x)&=&x^3-(8m^2+10 \kappa^2+3n^2)x^2+(16m^4+40 m^2
\kappa^2+24 \kappa^4+8 n^2 \kappa^2+3n^4)x\nn\\
&-&n^2(n^2-4m^2)(n^2-4m^2-2\kappa^2)\ .
\end{eqnarray}
 Following \cite{ft2},
we conclude that the stability condition
when all  $\omega$'s are  real is
\begin{equation}
f_{n}(0)=-n^2(n^2-4m^2)(n^2-4m^2-2\kappa^2)\leq 0\ .
\end{equation}
Here  $f_{0}(0)=f_{2m}(0)=0$, $f_{1}(0),...,f_{2m-1}(0)<0$, \
$f_{2m+1}(0)=-(2m+1)^2(4m+1)(4m+1-2\kappa^2)$. The stability
condition is then
\begin{equation}
\kappa^2< {2m+\ha }\ .
\end{equation}
Expressed  in terms of $\mathcal{S}$ it is
$\mathcal{S}\leq \mathcal{S}_{max}$, with
$
\mathcal{S}_{max}=\frac{4m+1}{8m^2}\sqrt{{(m+1)^2 - \ha }}
$.


\bigskip

The fermionic  fluctuations are described by the
  quadratic part of the \adss
superstring  Lagrangian evaluated on the  bosonic solution
\rf{crot}    (see \ci{MT,ft1,ft2,ft3} for details).
 \bea
 L_F=i\left(\eta^{ab}\d^{IJ}-\e^{ab}s^{IJ}\right)\;
 \bar{\th}^I\rho_aD_b\,\th^J\;\;,\;\;\ \ \ \rho_a\equiv \G_A
 e_a^A\ , \;\;\;\; e_a^A\equiv E_\m^A(\x)\pa_a \x^\m\ ,
 \eea
where $I,J=1,2,\;s^{IJ}=\mbox{diag}(1,-1),$ \ $ \rho_a$ are
projections of the ten-dimensional Dirac matrices and $\x^\m$ are
 the coordinates of the $AdS_5$  space for $\m=0,1,2,3,4$
 (i.e. $t,\r,\theta,\phi_1,\phi_2$) and
 the coordinates of $S^5$
for $\m=5,6,7,8,9$. The covariant derivative is given by
 \bea
 && D_a\th^I=\left(\d^{IJ}{\rm D}_a-\fr{i}2\e^{IJ}\G_*\rho_a\right)\th^J
 \;,\ \ \ \ \quad  \G_*\equiv i\G_{01234}\;,\;\ \G_*^2=1 \ ,
\eea
where ${\rm D}_a=\pa_a+\fr14\w_a^{AB}\G_{AB}, \ \  \w_a^{AB}\equiv
 \pa_a  \x^\mu  \w_\mu^{AB}$.
Fixing the $\k$-symmetry by the same condition as in \ci{ft3} \
 $\th^1=\th^2=\th $ \
one gets
 \bea
  L_F=-2i\bar{\th}D_F\th\;,\quad\quad  D_F=-\rho^a {\rm
  D}_a-\fr{i}2\e^{ab}\rho_a\G_*\rho_b\ .
 \eea
Explicitly, one finds
 \bea
D_F=&& \big(\k r_0 \G_0+w r_1 \G_3+w r_1 \G_4 \big) \nn\\
&&\times\ \big(\pa_\t-\fr{1}{\sqrt{2}}\k r_1\G_{10}-\fr{w r_0}{2\sqrt{2}}
     \G_{13}-\fr{w r_0}{2\sqrt{2}}\G_{14}-\fr{w }{2\sqrt{2}}
     \G_{23}+\fr{w }{2\sqrt{2}}\G_{24}\big) \nn\\
  &&  - \big(m r_1\G_3-m r_1\G_4\big)
    \big(\pa_\s-\fr{m r_0}{2\sqrt{2}}\G_{13}+\fr{m r_0}{2\sqrt{2}}\G_{14}
    -\fr{m}{2\sqrt{2}}\G_{23}-\fr{m}{2\sqrt{2}}\G_{24}\big)\nn\\
  && -\big(m\k r_0 r_1\G_3\G_0+2mw r_1^2\G_3\G_4-m\k r_0 r_1
   \G_4\G_0\big)\G_{01234}
 \eea
 Expanding the fields in terms of $e^{i \omega \tau + in \s}$
  can show that the $\det (D_F)=0$ condition leads to the
 characteristic  equation \rf{cup}. In the $4+6$
 dimensional representation of the  Dirac matrices
 one can use  the six dimensional $\Gamma$-matrices
 for the actual computation of the characteristic equation.
In \rf{cup}  there is  an  extra factor of 4 degeneracy making up the total
16 of fermionic  frequencies.

\renewcommand{\theequation}{C.\arabic{equation}}
 \setcounter{equation}{0}
  \section*{Appendix C: String  pulsating  in one plane
   }

Here we shall  review the pulsating $AdS_5$ solution of
\cite{minna} using the conformal gauge as in the case of the pulsating
solutions in $S^5$ discussed in  \ci{kt}. We shall start with
\rf{stringact} and consider the case where the string is pulsating
in one plane:
\begin{equation}
t=t(\tau)\ , \ \ \ \rho=\rho(\tau), \quad \theta=\frac{\pi}{2},
\quad \phi_{1}=m \sigma, \quad \phi_{2}=0
\end{equation}
Then the  only non-zero global $SO(2,4)$ charge is the energy
\begin{equation}
\mathcal{E}=\cosh^2 \rho\  \dot{t}\ .
\end{equation}
The  conformal gauge constraint gives  the equation of motion for
$\rho$
\begin{equation}
\dot{\rho}^2-\frac{\mathcal{E}^2}{\cosh^2 \rho}+m^2 \sinh^2
\rho=0\ ,
\end{equation}
which is the same as the $\S=0$ limit pf \rf{rho2}.
As in \ci{kt} the strategy is to   compute the oscillation  number
$\rN$ and then express the energy in terms of it.  By definition
(here $p_\rho = \dot \rho$)
\begin{equation}
\rN=\frac{\sqrt{\lambda}}{2\pi}\oint d\rho\
p_{\rho}=\frac{\sqrt{\lambda}}{\pi}\int_{0}^{\rho_{max}}d\rho
\sqrt{\frac{\mathcal{E}^2}{\cosh^2 \rho}-m^2 \sinh^2 \rho}
\end{equation}
Changing the  variable to $x=\sinh \rho$ we get
\begin{equation}
\rN=\frac{\sqrt{\lambda}}{\pi}\int_{0}^{\sqrt{R}}\frac{dx}{1+x^2}\sqrt{\mathcal{E}^2-m^2
x^2 (1+x^2)}\ , \ \ \ \ \ \ \ \ \
R=\frac{-m+\sqrt{m^2+4 \mathcal{E}^2}}{2m}\ .
\end{equation}
Then
\begin{equation}
\frac{\partial \rN}{\partial m}=-\frac{m \sqrt{\lambda}}{\pi}
\int_{0}^{\sqrt{R}} dx\frac{x^2}{\sqrt{\mathcal{E}^2-m^2 x^2
(1+x^2)}}
\end{equation}
 can be expressed in terms of the elliptic integrals
\begin{equation}
\frac{\partial \rN}{\partial m}=\frac{1}{2\pi} \sqrt{\frac{\l
a_{-}}{2m}}\ \frac{i a_{+}}{\mathcal{E}}\left(\textbf{K}\left[
\frac{a_{-}}{a_{+}}\right]-\textbf{E}\left[\frac{a_{-}}{a_{+}}\right]
\right)\ , \ \ \ \ \   a_{\pm}\equiv m \pm
\sqrt{m^2+4\mathcal{E}^2} \ .   \label{derivN}
\end{equation}
This representation is useful since it allows one to obtain
convergent  expansions for $\rN$ at large and small energies.

Expanding  in  large $\mathcal{E}$ and integrating  from $0$ to
$m$ we get
\begin{equation}
\rN(E,m)=\rN_{0}(E)+ { 1\ov 4} c_1
\lambda^{1/4}\sqrt{2mE}+O(E^{0})\ ,
\end{equation}
where \be \la{cc} c_1=\frac{4
\sqrt{2}}{\pi}\left(\textbf{E}[-1]-\textbf{K}[-1]\right)\approx
1.078\ .
\end{equation}
The constant of integration $\rN_{0}$ can be determined from the
$m=0$ integral\foot{The $m=0$ limit  corresponds to a massless
geodesic going from $\rho=0$ to $\rho=\infty$.}
\begin{equation}
\rN_{0}=\frac{\sqrt{\lambda}\mathcal{E}}{\pi}\int_{0}^{\infty}
\frac{dx}{1+x^2}=\ha E \ .
\end{equation}
Then inverting  the  relation between $\rN$ and $E$ we end up with
\begin{equation}
E=2\rN+ c_1  \lambda^{1/4} \sqrt{m\rN} +O(\rN^{0})\ ,
\label{minah}
\end{equation}
which  is the same expression as  found in \cite{minna}.

\bigskip

Expanding  (\ref{derivN})  at small  $\mathcal{E}$ gives
\begin{equation}
\frac{\partial \rN}{\partial
m}=\sqrt{\lambda}\left[\frac{\mathcal{E}^2}{4 m^2}-\frac{15
\mathcal{E}^{4}}{32
m^{4}}+O\left(\frac{\mathcal{E}^6}{m^{6}}\right)\right]\ .
\end{equation}
We observe that the right hand side is regular at $m\rightarrow
\infty$. Integrating  from $m$ to $\infty$ and setting
 $\rN(\mathcal{E},m=\infty)=0$ we get
\begin{equation}
\rN=\frac{E^2}{4 m}
\lambda^{-1/2}-\frac{45}{32}\frac{E^4}{m^3}\lambda^{-3/2}
+O\left(\frac{E^6}{m^5}\lambda^{-5/2}\right)\ ,
\end{equation}
and thus
\begin{equation}
E=2\lambda^{1/4} \sqrt{m\rN}+O(\rN^{3/2})\ .
\end{equation}
This is the flat-space dependence which is
 expected in the small-energy limit where the string is pulsating near the
 center of $AdS_5$.

\renewcommand{\theequation}{D.\arabic{equation}}
 \setcounter{equation}{0}
  \section*{Appendix D: String  rotating and pulsating \\
   in two planes
   }

Here we shall  derive the asymptotic expansions of the energy
of  the  general rotating and pulsating solution
discussed in section 3.

Let us first
 analyze the conditions when this  solution exists.
 We
need to have the expression inside the square root in \rf{ne}
 positive, i.e. where
 $f(y)\leq 0$ ($f(y)$ was defined in \rf{ko},\ $y=\sinh^2 \r$).
 The extrema of $f$,
 i.e. the solutions of
$f'(y)=0$ are
\begin{equation}
y_{\pm}=\frac{-m\pm \sqrt{m^2+3 (\mathcal{E}^2-4
\mathcal{S}^2)}}{3m}
\end{equation}
It follows that  $f(0)=4\mathcal{S}^2>0$, and
that $f(y)\rightarrow -\infty$
as $y\rightarrow -\infty$, with  $f(y)\rightarrow \infty$ as
$y\rightarrow \infty$. To have $f(y)\leq 0$ in a region inside
the allowed range of values
$y>0$ we need $y_{\pm}$ to be real. Since  $y_{-}<0$ to
have $f(y)\leq 0$ we need $y_{+}>0$. This requires
$\mathcal{E}^2\geq 4\mathcal{S}^2$. Then
$f(y_{-})>0$ and so  to have the
solution we need $f(y_{+})\leq 0$,
where
\begin{equation}
f(y_{+})=\frac{1}{27 m}\left[2m^3+9 m (8\mathcal{S}^2+
\mathcal{E}^2)
-[2m^2-6(4\mathcal{S}^2-
\mathcal{E}^2)]\sqrt{m^2+3(\mathcal{E}^2-4\mathcal{S}^2)}\right]
\end{equation}
From the form of $f(y_{+})$ we
conclude  that to have a solution we need
$g(z)\leq 0$ ($z\equiv \E^2$) for some $z$ within
the region $z>0$, where $g(z)$  was defined in \rf{g}.
We have  $g(0)>0$ and $g(z)\rightarrow -\infty$ as
$z\rightarrow \infty$, $g(z)\rightarrow \infty$ as $z\rightarrow
-\infty$. Then $g(z)=0$ has at least one real solution. We want
$g(z)\leq 0$ in a region within $z>0$. Therefore, we need
$z\equiv  \mathcal{E}^2 \geq h(m,\mathcal{S})$
where  $h(m,\mathcal{S}$ is
the maximal root  of $g(z)$ (if the maximal root is
negative we just need $z\geq0$).
 In addition, we have to satisfy
  $\mathcal{E}^2\geq 4\mathcal{S}^2$, so we conclude
that the solution exists if
\begin{equation}
\mathcal{E}^2 \geq \max \left\{4 \mathcal{S}^2,
h(m,\mathcal{S})\right\}
\end{equation}
Note that for small or large  $\mathcal{S}$ we have
$h(m,\mathcal{S})> 4\mathcal{S}^2.$

To study the small energy behavior
$\mathcal{E}\rightarrow 0$, we thus need also
$\mathcal{S}\rightarrow
0. $ In this limit
$h(m,\mathcal{S})=4m\mathcal{S}+8\mathcal{S}^2+O(\mathcal{S}^3)$,
so we are to require
$\mathcal{E}^2\geq 4m \mathcal{S}$.
The case of  $\mathcal{E}^2= 4m \mathcal{S}$
  corresponds to $\rho$
constant, i.e. to no pulsation or  $\mathcal{N}=0$, when, as in
\rf{sma}, we get
\begin{equation}
\mathcal{E}=2\sqrt{m\mathcal{S}}+{2
m^{-1/2} \mathcal{S}^{3/2}}+O(\mathcal{S}^{5/2})\ .
\end{equation}
More generally, one finds from \rf{osc} that in the limit of
small $\E$ and $\S$
\begin{equation}
E=2\lambda^{1/4}\sqrt{m(S+\rN)}+...\ .
\end{equation}
Again, this  is the flat-space behavior, which is expected in this case
since the string is located near the origin of
$AdS_5$.

\bigskip

In the large energy limit  we can expand
$h(m,\mathcal{S})$ at large $\mathcal{S}$. In the case of the
equality
$\mathcal{E}^2=h(m,\mathcal{S})$, which again corresponds to
constant $\rho$, i.e.  $\mathcal{N}=0$,
 we reproduce the asymptotics of the rigid string case  \rf{la}
\begin{equation}
\mathcal{E}=2\mathcal{S}+\frac{3}{4}(4m^2)^{1/3}\mathcal{S}^{1/3}+
O(\mathcal{S}^{-1/3})
\end{equation}
Let us now  analyze the large energy behavior
 starting from
(\ref{osc}) or
\begin{equation}
\frac{\partial \rN}{\partial m}=-\frac{m \sqrt{\lambda}}{\pi}
\int_{\sqrt{R}_{2}}^{\sqrt{R}_{3}}dx
\frac{x^2}{\sqrt{\mathcal{E}^2-\frac{4 \mathcal{S}^2
(1+x^2)}{x^2}-m^2 x^2 (1+x^2)}}\ .
\end{equation}
This integral can be expressed in terms of the elliptic integrals
$\textbf{E}$,$\textbf{F}$
\begin{equation}
\frac{\partial \rN}{\partial m}=\frac{m \sqrt{\lambda}}{\pi \sqrt{R_{3}-R_{1}}}\left(
(R_{1}-R_{3})\textbf{E}\left[\frac{R_{2}-R_{3}}{R_{1}-R_{3}}\right]-R_{1}
\textbf{F}\left[\frac{R_{2}-R_{3}}{R_{1}-R_{3}}\right]\right)
\ .
\label{partn}
\end{equation}
The large energy behavior of the roots $R_{1}$,$R_{2}$,$R_{3}$ of $f(y)$ in \rf{ko} is
\begin{equation}
R_{1,3}=\mp \frac{\mathcal{E}}{m}-\frac{1}{2}\mp \frac{m^2-16
\mathcal{S}^2}{8 m \mathcal{E}}+O(\mathcal{E}^{-2}), \quad\ \ \ \
\
R_{2}=\frac{4 \mathcal{S}^2}{\mathcal{E}^2}+O(\mathcal{E}^{-4})\ .
\end{equation}
Expanding for large energy $\mathcal{E}$ we obtain
\begin{eqnarray}
&&\frac{\partial \rN}{\partial
m}=\frac{\sqrt{\lambda}}{\pi\sqrt{2}}\bigg[  k_1 \sqrt{m \mathcal{E}}+
\frac{ k_2   m^{3/2}   }{4\sqrt{2}\sqrt{\mathcal{E}}}
- \frac{k_1 \sqrt{m}}{32 \mathcal{E}^{3/2} }(3m^2-32
\mathcal{S}^2) +O(\mathcal{E}^{-5/2})\bigg]\ ,
\end{eqnarray}
\be
k_1\equiv
\textbf{K}\left[\frac{1}{2}\right]
-2\textbf{E}\left[\frac{1}{2}\right]  \ , \ \ \ \ \ \ \ \ \ \ \
k_2 \equiv \textbf{K}\left[\frac{1}{2}\right]
\ee
Integrating over $m$ we obtain $\rN$
\begin{eqnarray}
\rN(\mathcal{E},\mathcal{S})=\rN_{0}(\mathcal{E},\mathcal{S})
+  \frac{\sqrt{2\lambda}m^{3/2}  }{\pi
}\bigg[
\frac{2 k_1 \sqrt{\mathcal{E}}}{9}+
\frac{ k_2 m }{20 \sqrt{\mathcal{E}}}
- \frac{k_1   \left(\frac{3}{7}m^{2}-\frac{32}{3}
\mathcal{S}^2\right) }{32 {\mathcal{E}^{3/2}} } \bigg]
  + ...
\end{eqnarray}
The integration constant
$\rN_{0}(\mathcal{E},\mathcal{S})$
can be determined  from the integral (\ref{osc})
for $m=0$
\begin{equation}
\rN_{0}=\frac{\sqrt{\lambda}}{\pi}\int_{\sqrt{R}}^{\infty}\frac{dx}{1+x^2}\sqrt{\mathcal{E}^2-\frac{4\mathcal{S}^2
(1+x^2)}{x^2}}\ , \ \ \ \ \ \ \ \
R^2=\frac{4 \mathcal{S}^2}{\mathcal{E}^2-4 \mathcal{S}^2}\ .
\end{equation}
We find
\begin{equation}
\rN_{0}=\frac{2
\mathcal{S}\sqrt{\lambda}}{\pi}\int_{1}^{\infty}\frac{dx}{1+R^2
x^2}\sqrt{1-\frac{1}{x^2}}= \ha \sql ( \E -2 \S)\ .
\end{equation}
Note  that in the limit of $m\rightarrow 0$
(or zero tension)
 the  solution is the same as  for
 a point particle\foot{Note that the solution \rf{a}
depends on  $\s$ only in combination with $m$.}
moving along  a massless  geodesic in $AdS_5$ from some
finite $\rho$ to $\rho=\infty$. We may thus  interpret the
string  solution with  large
energy (implying large spin $\mathcal{S}$ or large oscillation
number $\mathcal{N}$) as being  ``fast'' (cf. \ci{mik12}).
For $\E\sim 1$  which we considered in section 3,
the string is not fast:
 in this case the  $m\to 0$ limit
corresponds  to the radial coordinate
$\rho$ changing between two finite values.

The resulting expression for the  oscillation number is
\begin{equation}
\rN=\frac{E}{2}-S+a_{1}
\lambda^{1/4} \sqrt{E}+a_{2}
\lambda^{3/4} \frac{1}{\sqrt{E}}+O(E^{-3/2})\ ,
\label{ns}
\end{equation}
where
\begin{equation}
a_{1}=\frac{\sqrt{2} k_1 }{3\pi} m^{3/2}= -0.127m^{3/2}
\  ,\ \ \ \ \ \ \ \ \ \ \  \quad a_{2}=\frac{k_2}{10 \sqrt{2}
\pi}m^{5/2} =0.042 m^{5/2}
\end{equation}
Expressing energy  in terms of $\rN$ and $S$
then gives \rf{smallS}.
 Since we expanded in  large energy while
keeping $\S$ fixed, this expression is valid for large $\E$ and
$\N$, but not large $\S$.\foot{
Note that in
 the limit $S\rightarrow 0$ we obtain
 from \rf{smallS} a different expression for
the energy then found in the absence of rotation in \rf{minah},
in spite  of the fact that setting $\S=0$ in \rf{rho2}
gives the same equation as in the case of string pulsating in one
plane.
 This discontinuity is explained by the
fact that taking the limit
$\mathcal{S}\neq 0$ changes the behavior of the
potential at $\rho=0.$ With nonzero $\mathcal{S}$ the potential
blows up at $\rho=0$, while with $\mathcal{S}=0$ it is finite at
$\rho=0$.}


\bigskip

Let us now  find the energy  for large
$\mathcal{S}$ and small $\mathcal{N}$.
Let us start with  (\ref{partn}) and expand now at
large  $\mathcal{S}$ and
large  $\mathcal{E}$, setting
$\mathcal{S}=u\mathcal{E}$, with $u$ fixed.
For the solution to
exist we need to satisfy (\ref{cond}) for large $\mathcal{S}$,
which implies
 $u\leq \frac{1}{2}.$ Expanding  in  large energy
$\mathcal{E}$ but  with $u$ fixed we obtain
\begin{equation}
\rN=\frac{E}{2}-S+a_{1} \lambda^{1/4}
\sqrt{E}\left(1-4\frac{S^2}{E^2}\right)^{1/4}+
a_{2}\lambda^{3/4}\frac{1}{\sqrt{E}}
\frac{1-8\frac{S^2}{E^2}}{\left(1-4\frac{S^2}{E^2}\right)^{5/4}}
+O(E^{-3/2})
\end{equation}
This reduces to  (\ref{ns})
if we expand in  small $u$.
In the limit  $u\rightarrow
\frac{1}{2}$ we  get
\begin{equation}
\rN=\frac{E}{2}-S+a_{1} \lambda^{1/4}
\sqrt{E}\left(\frac{E}{S}-2\right)^{1/4}-a_{2}\lambda^{3/4}\frac{1}{\sqrt{E}}
{\left(\frac{E}{S}-2\right)^{-5/4}}+O(E^{-3/2})\ ,
\end{equation}
i.e.
\begin{equation}
E=2S+ b_0 \rN + b_{1}\lambda^{1/3}S^{1/3}+b_{2}
{\lambda^{3/2}}S^{-1/3}+...
\end{equation}
where  for $m=1$  we have  $b_{0}=1.835$, $b_{1}=0.454$.
For $\rN\to 0$, this is different from  the expression \rf{la}
in the
case of the
rigid  rotating string, which for $m=1$ gives
$E=2S+1.19 (\lambda S)^{1/3}+ ....$ While the $S$ dependence
 is the same the
coefficients are different. This is
an indication of the same discontinuity  we mentioned above
in the case of large energy and
 large $\rN$ and the limit
$S\rightarrow 0.$
In this sense  there is a symmetry
between the $\N$ and $\S$ dependence of the energy.

\bigskip




\end{document}